\documentclass[conference]{IEEEtran}
\IEEEoverridecommandlockouts
\usepackage{cite}
\usepackage{amsmath,amssymb,amsfonts}
\usepackage{algorithmic}
\usepackage{graphicx}
\usepackage{textcomp}
\usepackage{xcolor}
\def\BibTeX{{\rm B\kern-.05em{\sc i\kern-.025em b}\kern-.08em
    T\kern-.1667em\lower.7ex\hbox{E}\kern-.125emX}}

\begin{document}
\title{Link-Layer Rate of NOMA with Finite Blocklength for Low-Latency Communications }
\author{\IEEEauthorblockN{Muhammad Amjad$^\dagger$, Leila Musavian$^\dagger$, and Sonia A\"{i}ssa$^\ddagger$}
\IEEEauthorblockA{$^\dagger$School of Computer Science
and Electronic Engineering (CSEE),  University of Essex, UK.\\
$^\ddagger$Institut National de la Recherche Scientifique (INRS-EMT), University of Quebec, Montreal, QC, Canada.\\
Email: \{m.amjad, leila.musavian\}@essex.ac.uk, aissa@emt.inrs.ca}}

\maketitle

\begin{abstract}
Non-orthogonal multiple access (NOMA) in conjunction with finite blocklength (short packet) communications is regarded as an enabler for ultra-reliable and low-latency communications (URLLC). In this paper, we investigate the link-layer rate, i.e., the effective capacity, of a two-user NOMA in finite blocklength regime. The delay performance of the NOMA users is analyzed by taking into consideration the queueing delay violation probability and the transmission error probability. We further provide closed-form expressions for the individual effective capacity of the NOMA users in Rayleigh fading environment. Through simulations, we investigate the impact of the transmit signal-to-noise ratio and the delay exponent on the achievable effective capacity and the queueing delay violation probability of the NOMA weak and strong users. In particular, results show that when using short packet communications, the queueing delay violation probability cannot be improved below a threshold.
\end{abstract}
\section{Introduction}
\label{sec:in}
The ballooning growth of new applications, such as tactile Internet, massive sensing, holographic teleportation, and autonomous vehicles, poses serious challenges in terms of provisioning the spectrum efficiency, higher reliability, and lower latency \cite{R136}. The 5th generation (5G) and beyond 5G (B5G) promise seamless connectivity, enhanced capacity,  higher reliability, and low-latency, through its use-cases of ultra-reliable and low-latency communications (URLLC), and massive machine-type communications (mMTC). To meet the target objectives, research on key enabling technologies has proliferated, in particular on non-orthogonal multiple access (NOMA) as a promising access technique for 5G and B5G. Recently, using NOMA in conjunction with short packet communications has been explored to achieve low latency and high reliability \cite{R140}.

Though the shortening of packets can be straightforwardly convincing as a mean to achieve URLLC, it poses serious challenges such as capacity penalty \cite{R139}. As known, traditional wireless systems are designed based on the Shannon theory, where long packets are considered in the communications. The Shannon limit, however, is a loose upper bound for the performance of systems with short packet communications \cite{R153}. Serval research efforts have been done to evaluate the performance of communication systems with finite blocklength, and advance the theory of short packet communications in general \cite{R139,R140}. In particular, the pioneering work in \cite{R139} assessed the service rate of short packet communications for additive white Gaussian noise (AWGN) channels.\footnote{The terms finite blocklength and short packet communications will be used interchangeably throughout the paper.} The trade-off between reliability, latency, and throughput of mission-critical MTC subject to Rayleigh block-fading was investigated later in \cite{R155}, which derived the maximum achievable coding rate in a finite blocklength regime.

Recently, the suitability of finite blocklength communications for low latency or URLLC received special attention by researchers \cite{R157}. The latency performance of a point-to-point communications system with finite blocklength was also studied by using the effective capacity framework. Effective capacity is the dual concept of effective bandwidth, and defines the maximum arrival/source rate for a given service rate while satisfying a certain delay quality-of-service (QoS) exponent (see \cite{R176,R158} and references therein). The achievable effective capacity of MTC in finite blocklength regime was investigated in \cite{R156}, which provided an upper bound on the effective capacity, and an optimal power allocation policy that provides the trade-off between the delay performance and power consumption. A closed-form expression for the effective capacity in finite blocklength regime was presented, and its accuracy was confirmed using Monte-Carlo simulations. In \cite{R154}, using an effective capacity model in finite blocklength regime, a trade-off between the coding rate, throughput, blocklength, and queuing constraint, was investigated. Therein, various transmission strategies, i.e., fixed-rate, variable rate, and variable-power transmissions, were studied for short packet communications using the effective capacity metric.

Considering a NOMA architecture, the authors in \cite{R159} investigated the effective capacity of a two-user system in conventional long-packet communications regime. The total link-layer rate of a two-user NOMA was also estimated, and compared with its OMA counterpart. This study revealed that the effective capacity of NOMA outperforms the one for orthogonal multiple access (OMA) at high signal-to-noise ratios (SNRs). Later, considering short-packet communications, the effective-bandwidth of NOMA was investigated in \cite{R160}, which provided the required SNR for a given transmission error probability and delay exponent. However, the detailed latency performance of NOMA, i.e., the impact of delay exponent on the queueing delay violation probability, and the closed-form expression of the achievable EC, are yet to be performed and analyzed.

In this paper, we investigate the achievable link-layer rate of a two-user NOMA   with short-packet communications.  Specifically, we formulate the effective capacity of the strong and weak users under heterogeneous delay QoS requirements. The overall reliability, which is the combination of the transmission error probability and the queueing delay violation probability, is investigated. We also derive closed-form expressions for the individual effective capacity of the two NOMA users. Through simulations, we investigate the impact of transmit SNR and the delay exponent on the achieved effective capacity and on the queueing delay violation probability. In particular, we show that when the delay exponent increases, the queueing delay violation probability cannot be improved below a certain value due to the dominance of the transmission error probability.

The remainder of the paper is organized as follows. Section \ref{sec:system-model} describes the system model. Section \ref{sec:ec-noma-short} discusses the effective capacity of the NOMA system in finite blocklength regime. Numerical results are provided in Section \ref{sec:numerical-results}, followed by conclusions in Section \ref{sec:conclusion}.
\section{System Model}
\label{sec:system-model}
We consider a downlink power-domain NOMA operation with short-packet communications. The system consists of a base station (BS) communicating and $V$ single-antenna users, and we consider block fading for any channel between the BS and the individual user such that the channel power gain remains constant during one fading block and changes from one fading block to another. Without loss of generality, we assume that the users' channels are sorted such that $\left | h_{1}(\tau)  \right |^2\leq \left | h_{2} (\tau)   \right |^2\leq...\leq\left | h_{V}(\tau)   \right |^2$, where $h_j(\tau)$ is the channel gain between the BS and the $j^{\rm th}$ user at time $\tau$. We assume that two users, out of the set of $V$ users, and  referred to by $v_i$, $i\in\{t,u\}$, such that $1\leq t < u \leq V$, are scheduled for service in the shared resource block. The two-user NOMA is considered due to its importance as part of 3GPP-LTE Advance group \cite{R175}.

\begin{figure}[th]
\centering
  \includegraphics[width=0.8\linewidth,height=5.5cm]{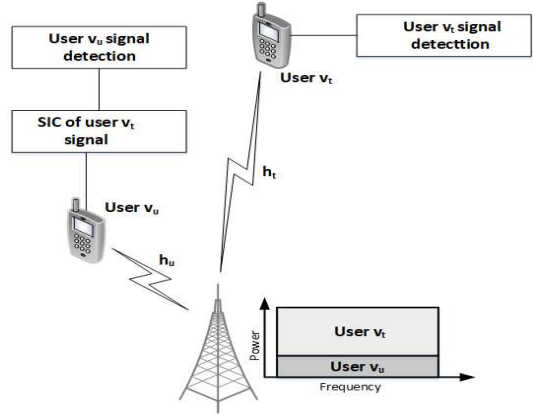}
\caption{Two-user NOMA basic architecture.}
\label{fig:sys1}
\end{figure}

\begin{figure*}
\centering
\includegraphics[width=14cm,height=8cm]{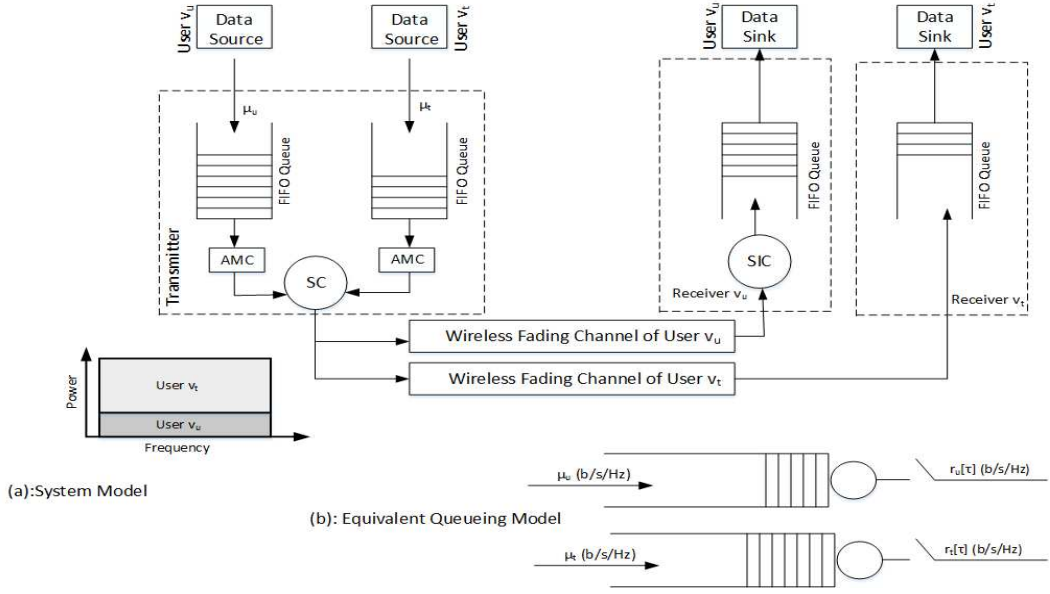}
\caption{Two-user NOMA system model.}
\label{fig:sys2}
\end{figure*}

Figures \ref{fig:sys1} and \ref{fig:sys2} show the basic operation of the two-user NOMA-based short-packet communication system, and the corresponding queuing model with the effective capacity concept. According to the NOMA principle, the BS broadcasts a combined message of both users, according to $\sum_{i\in\{t,u\}}\sqrt{\alpha_{i}P}s_{i}(\tau)$, where $s_{i}(\tau)$  is the message of user $v_i$ at time $\tau$, $\alpha_{i}$ is the power allocation coefficient of user $v_i$ such that $\alpha_t \textgreater \alpha_u$, and $P$ is the BS power. As mentioned, the users are classified based on their channel conditions such that $\left | h_{u}(\tau)  \right |^2\geq \left | h_{t} (\tau)   \right |^2$, where $h_u(\tau)$ and $h_t(\tau)$ are the channel coefficients between the BS and users $v_u$ and $v_t$ at time $\tau$, respectively.


Following the NOMA operation, the received signal at $v_i$ is given by $y_i=h_i\sum_{i\in\{t,u\}} \sqrt{\alpha_{i}P}s_i  + m_i$,\footnote{As the channel coefficients are assumed stationary and ergodic random processes, the time index $\tau$ is omitted hereafter.} where $m_i$ is the additive white Gaussian noise (AWGN).  After receiving the signal from the BS, the weak user $v_t$ decodes $s_t$ by considering $s_u$ as interference. Therefore, the resulting signal-to-interference-plus-noise ratio (SINR) at $v_t$  is given by
\begin{equation}
\label{eq:snr_t}
 \gamma_{t}=\frac{\alpha_{t}\left | h_{t} \right |^2 }{\alpha_{u}\left | h_{t} \right |^2+\frac{1}{\rho }},
\end{equation}
where $\rho$ is the transmit SNR, i.e., $\rho=\frac{P}{N_{o}B}$, with ${N_{o}B}$ denoting the noise power. The strong user, however, performs successive interference cancellation (SIC), and removes $s_t$ from its received message first. The received SNR at $v_u$ can hence be found as
\begin{equation}
\label{eq:snr_u}
 \gamma_{u}=\alpha_{u}\rho\left | h_u \right |^2 .
\end{equation}
In this work, the channel gains are modeled as  Rayleigh fading with unit variance. Since in NOMA the users are sorted based on the ordered channel gains, the statistics of the ordered channel power gains follows the order statistics \cite{R163}. The probability density functions (PDFs) of the ordered $\gamma_t$ and $\gamma_u$ \cite{R159} are given by 
\begin{equation}
\label{eq:pdfclosed_t}
f_{(t)}\left ( \gamma_{t} \right )=\xi_{t}f\left ( \gamma_{t} \right )F\left ( \gamma_{t} \right )^{t-1}\left (1-F\left ( \gamma_{t} \right )       \right )^{V-t},
\end{equation}
\noindent and
\begin{equation}
\label{eq:pdfclosed_u}
\hspace{0.8cm} f_{(u)}\left ( \gamma_{u} \right )=\xi_{u}f\left ( \gamma_{u} \right )F\left ( \gamma_{u} \right )^{u-1}\left (1-F\left ( \gamma_{u} \right )       \right )^{V-u},
\end{equation}
respectively, where $f_{(t)}\left ( \gamma_{t} \right )$ and $f_{(u)}\left ( \gamma_{u} \right )$ are the PDFs of $\gamma_{t}$ and $\gamma_u$, $F(\gamma_t)$ and $F(\gamma_t)$ are the cumulative distribution functions (CDFs) of $\gamma_{t}$  and $\gamma_u$, respectively, $\xi_{t}=\frac{1}{B\left ( t,V-t+1 \right )}$, $\xi_{u}=\frac{1}{B\left ( u,V-u+1 \right )}$, and $B(a,b)$ is the Beta function.

As we are considering the transmission of short packets in our two-user NOMA operation, the achievable rate is not only the function of SNR (or SINR), but also the transmission error probability ($\epsilon$), and the blocklength ($n$). Hence, the achievable rates of the users   can expressed as \cite{R140}
\begin{align}
\label{eq:rate_i}
\begin{split}
r_{i} = {\rm ln}\left ( 1+\gamma_{i} \right )  -  \sqrt{\frac{\delta_{i}}{n}}Q^{-1}{(\varepsilon_{i})},
\end{split}
\end{align}
where  $r_{i}$ is the achievable rate of user $v_i$, $i\in\{t,u\}$, $\delta_i$ is the channel dispersion which can be approximated by $\delta_{i}=\sqrt{1-(1+\gamma_{i} )^{-2}}$,   $Q^{-1}(.)$ is the inverse of Gaussian Q-function, $\epsilon_i$  is the transmission error probability corresponding to user $v_i$.
\section{Effective Capacity of Downlink NOMA with Finite Blocklength}
\label{sec:ec-noma-short}
In this section, we aim to derive the achievable effective capacity the NOMA system with finite blocklength communications described above. To proceed, required concepts are recalled.

Motivated by the theory of effective bandwidth, authors in \cite{R147} introduced the effective capacity metric as the dual concept of effective bandwidth. Effective capacity is the maximum arrival (source) rate that the channel can support (service rate) with certain delay requirements. Here, we assume that the transmission technique from the BS to the NOMA users satisfy the delay constraint. We also assume that, at the transmit buffer, the probability of queue length exceeding a certain threshold $x$, exponentially decays as a function of $x$ \cite{R161,R162}, where $\theta_i$ is the delay QoS exponent and can be approximated using
\begin{equation}
\label{eq:ec_theta1}
-\lim_{x\rightarrow \infty}\frac{{\rm{ln}}\left ({\rm{Pr}}\left \{Q_i(\infty)> x  \right \}  \right )}{x}=\theta_i,
\end{equation}
where $Q_i(\infty)$ is the steady-state of the transmit buffer corresponding to the data of user $v_i$, $i\in\{t,u\}$, and ${\rm{Pr}\{a\textgreater b\}}$ is the probability that $a \textgreater b$ holds.

Using (\ref{eq:ec_theta1}), the queuing delay violation probability can be approximated as
\begin{equation}
 \label{eq:qdvp}
 {\rm{Pr}}\left \{  D_i>D_{{i}}^{\rm{max}}\right \}\approx {\rm{Pr}}\left \{  Q_i(\infty)>0\right \}e^{-\theta_i \mu_i D_{{i}}^{\rm{max}}},
\end{equation}
where, w.r.t. $v_i$, ${\rm{Pr}}\left \{ Q_i(\infty)>0 \right \}$ is the probability of non-empty buffer, $D_{{i}}^{\rm{max}}$ is the maximum delay, and  $\mu_i$ is the maximum arrival rate \cite{R147}. The more stringent delay requirements can be represented with larger values for $\theta_i$,  while smaller values of $\theta_i$ indicate less stringent delay requirements.

By taking into account the QoS exponent $\theta_i$, the maximum arrival rate, i.e., the achievable effective capacity of user $v_i$, in b/s/Hz, can be formulated as
\begin{equation}
\label{eq:ec_i}
C_{e}^{i}=-\frac{1}{\theta_{i}{n}}{\rm{ln}}\left (\mathbb{E} \left [ \epsilon+\left ( 1-\epsilon \right ) e^{-\theta_{i}{n}{r_i}}  \right ]  \right ),
\end{equation}
where $\mathbb{E}[.]$ is the expectation operator. As we are using finite blocklength, the service rate with finite blocklength for $v_i$ user, i.e., $r_i$ in the above expression, can be replaced with (\ref{eq:rate_i}).

The effective capacity for the weak user and the strong user of the NOMA system, in b/s/Hz, can now be approximated as
\begin{equation}
 \label{eq:ec_i}
C_{e}^{i}=-\frac{1}{\theta_{i}{n}}{\rm{ln}}\left (\mathbb{E} \left [\epsilon+\left ( 1-\epsilon \right )
\left [ 1+\gamma_{i} \right ]^{2\zeta_{i}} e^{\beta_{i}\delta_{i}}  \right ]  \right ),
\end{equation}
where  $\zeta_{i}=-\frac{\theta_{i}{n}}{2{\rm{ln}2}}$,  $\beta_{i}=\theta\sqrt{n}Q^{-1}{(\varepsilon_{i})}$,  $\delta_{i}=\sqrt{1-(1+\gamma_{i} )^{-2}}$, and $i\in\{t,u\}$.

Using (\ref{eq:snr_t}) and  (\ref{eq:snr_u}), the effective capacity of the weak user and the strong user of NOMA with finite blocklength regime can finally be found, in the unit of b/s/Hz, as
\begin{equation}
\begin{split}
 \label{eq:ec_t3}
C_{e}^{t}=-\frac{1}{\theta_{t}{n}}{\rm{ln}}\Bigg(\mathbb{E} \Bigg[\epsilon+\left( 1-\epsilon \right)   &
\left(1+\frac{\alpha_{t}\left | h_{t} \right |^2 }{\alpha_{u}\left | h_{t} \right |^2+\frac{1}{\rho }}  \right)^{2\zeta_{t}} \\ & e^{\beta_{t}\delta_{t}}  \Bigg]  \Bigg ),
\end{split}
\end{equation}
\noindent and
\begin{equation}
\begin{split}
 \label{eq:ec_u3}
C_{e}^{u}=-\frac{1}{\theta_{u}{n}}{\rm{ln}} \Bigg (\mathbb{E} \Bigg [\epsilon+\left ( 1-\epsilon \right )   &
\Big ( 1+\alpha_{u}\rho\left | h_u \right |^2  \Big )^{2\zeta_{u}} \\ &   e^{\beta_{u}\delta_{u}}  \Bigg ]  \Bigg ),
\end{split}
\end{equation}
respectively.
\begin{figure*}
\begin{align}
&\begin{aligned}
 \begin{split}
 \label{eq:ec_closed_t}
 C_{e}^{t}&\approx -\frac{1}{\theta_{t}n}{\rm{ln}} \left [ \epsilon+\left ( 1-\epsilon \right )    \left \{ \left (\frac{\alpha_{u}^{-2\zeta_{t}}\xi_{t} }
  {\rho}\left ( \sum_{r=0}^{t-1}\binom{t-1}{r}(-1)^{r}\frac{1}{\eta_{t}\alpha_{u}}   +\frac{\theta_{t}n(\alpha_{u}-1)}    {\alpha_{u}{\rm{ln}2}} \sum_{r=0}^{t-1}\binom{t-1}{r}(-1)^{r}
     e^{\eta_{t}}E_i (-\eta_t) \right. \right.\right.\right.  \\
         &\left. \left.   +\sum_{s=2}^{\infty}  \binom{2\zeta_{t}}{s}\left (\frac{\alpha_{u}-1}{\alpha_{u}} \right)^s \sum_{r=0}^{t-1}\binom{t-1}{r}(-1)^{r}\left (\frac{\sum_{r=1}^{s-1} \frac{(r-1)!}{\alpha_{u }^{-r}} (-\alpha_u \eta_{t})^{s-r-1}    }{(s-1)!}-    \frac{(-\alpha_u \eta_{t})^{s-1}}{(s-1)!}        e^{\eta_{t}}    E_{i}(-\eta_{t})
  \right )     \right )  \right )\left ( K_t+1 \right )\\
  &   - \left ( \left (  \frac{\alpha_{u}^{-(2\zeta_{t}-2)}\xi_{t} }
  {\rho}\left ( \sum_{r=0}^{t-1}\binom{t-1}{r}(-1)^{r}\frac{1}{\eta_{t}\alpha_{u}}   +\frac{\theta_{t}n(\alpha_{u}-1)}    {\alpha_{u}{\rm{ln}2}} \sum_{r=0}^{t-1}\binom{t-1}{r}(-1)^{r}
     e^{\eta_{t}}E_i (-\eta_t)+\sum_{s=2}^{\infty}
          \binom{2\zeta_{t}-2}{s}\left (\frac{\alpha_{u}-1}{\alpha_{u}} \right)^s  \right. \right.\right.  \\
          &\left. \left. \left. \left. \left.  \times \sum_{r=0}^{t-1} \binom{t-1}{r}(-1)^{r}\left (\frac{\sum_{r=1}^{s-1} \frac{(r-1)!}{\alpha_{u }^{-r}} (-\alpha_u \eta_{t})^{s-r-1}    }{(s-1)!}-    \frac{(-\alpha_u \eta_{t})^{s-1}}{(s-1)!}        e^{\eta_{t}}    E_{i}(-\eta_{t})
  \right )     \right ) \right ) \left ( K_t-\frac{\beta_{t}}{2} \right ) \right ) \right \}   \right ].
\end{split}
\end{aligned}\\
&\begin{aligned}
 \begin{split}
\label{eq:ec_closed_u}
C_{e}^{u}\approx -\frac{1}{\theta_{u}{n}}{\rm{ln}}\left ( \epsilon+\left ( 1-\epsilon \right )\left (\frac{\xi_{u}}{\rho\alpha_{u}} \sum_{i=0}^{u-1}\binom{u-1}{i}(-1)^{i}  \left \{{\rm{H}}(1,2+2\zeta_{u},\eta_{u})({\rm{K_u+1}} )- {\rm{H}}(1,2\zeta_{u},\eta_{u})\left({\rm{K_u}}-\frac{\beta_{u}}{2} \right)  \right \} \right )   \right ).
\end{split}
\end{aligned}
\end{align}
\rule{18cm}{0.02cm}
\end{figure*}

To further simplify and to obtain the closed-form expressions for the above derived individual effective capacity of the two users, we use the Maclaurin series \cite{R172}  for the expression $e^{\beta_{u}\delta_{u}}$. This expression is simplified by the approximation according to $e^{\beta_{i}\delta_{i}} \approx 1+  \beta_{i}\delta_{i}  +\frac{\left (\beta_{i}\delta_{i}  \right )^2}{2}$. Also, the term $\delta_{i}=\sqrt{1-(1+\gamma_{i} )^{-2}}$ is  further simplified by using the  Laurent's expansion \cite{R170}, such that $\delta_{i} \approx 1-\frac{1}{2}(1+\gamma_{i} )^{-2}$. These simplifications pave the way for finding closed-form expressions for the individual EC each NOMA user. We note that these simplifications introduce some mismatch in verifying the close-form and Monte-carlo simulations, as detailed later in Section \ref{sec:numerical-results}.

Making use of the PDFs shown in (\ref{eq:pdfclosed_t}) and (\ref{eq:pdfclosed_u}), the closed-form expressions for the effective capacity of the weak and strong users in finite blocklength regime are obtained, as shown in (\ref{eq:ec_closed_t}) and (\ref{eq:ec_closed_u}), where $\eta_{u}=\frac{V-u+1+i}{\rho\alpha_u}$, $\eta_{t}=\frac{V-t+1+r}{\rho\alpha_u}$, $K_t=\frac{\beta_{t}^{2}}{2}+\beta_t$,   $K_u=\frac{\beta_{u}^{2}}{2}+\beta_u$, $E_i(.)$ is the exponential integral given by $E_i(x)=-\int_{-x}^{\infty}\frac{e^{-t}}{t}dt$ \cite{R171}, and $H(a,b,z)$ is the confluent Hypergeometric function of the second kind \cite{R171}, i.e.,
\begin{align}
 \begin{split}
 \label{eq:app_u5}
{\rm{H}}\left ( a,b,z \right )=\frac{1}{\Gamma\left ( a \right )} \int_{0}^{\infty}e^{-zy}y^{a-1}\left ( 1+y \right )^{b-a-1} d_{y} \\ \ \rm{for} \ \rm{Re } \ (a),\ \rm{Re} \ (z)>0,
\end{split}
\end{align}
in which ${\Gamma\left ( a \right )}$ denotes the Gamma function.
\section{Numerical Results}
\label{sec:numerical-results}
In this section, we provide simulation results of the proposed two-user NOMA  in finite blocklength regime. The total number of users is taken as $V=10$. We have set $u=8$, and $t=2$ as the indices of the strong and weak users, respectively, let $\alpha_u=0.2$ and $\alpha_t=0.8$ be the respective power coefficients in the resource allocation at the BS.

\begin{figure}[ht]
\centering
 \includegraphics[width=0.9\linewidth]{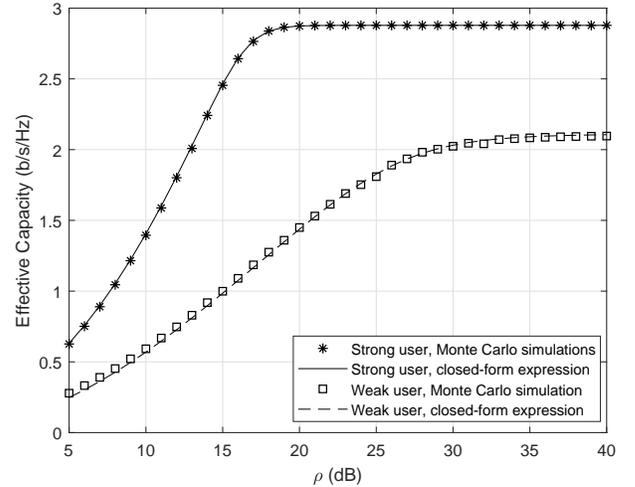}
\caption{Effective capacity of NOMA weak user and strong user versus transmit SNR, with $\theta=0.01$, $n=300$, and $\epsilon=10^{-5}$.}
\label{fig:closed-monte}
\end{figure}

Fig. \ref{fig:closed-monte} shows the curves for the achievable effective capacity of the paired strong and weak users, in b/s/Hz, versus the transmit SNR, $\rho$ in dB. The delay exponent is set as $\theta=0.01$, the blocklength $n=300$, and the transmission error probability is $\epsilon=10^{-5}$. This figure also shows the comparison of the closed-form expressions with Monte-Carlo simulations, and confirms the accuracy of the proposed closed-form expressions. The small mismatch between the analytical results and the Monte-carlo simulations corresponding to the weak user is due to the usage of the approximations for deriving the closed form expressions. The Maclaurin series and the Laurent expansion are used for the approximation of two factors, namely, $e^{\beta_{i}\delta_{i}}$ and $\delta_{i}$, as detailed in Section \ref{sec:ec-noma-short}.

\begin{figure}[ht]
\centering
 \includegraphics[width=0.9\linewidth]{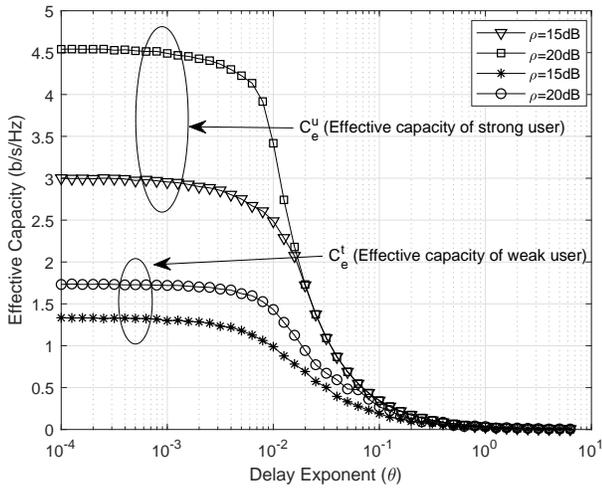}
\caption{Effective capacity of NOMA weak user versus delay exponent $\theta$, with $n=400$ and  $\epsilon=10^{-6}$.    }
\label{fig:ec-theta-weak-strong}
\end{figure}

Fig. \ref{fig:ec-theta-weak-strong} shows the plots of achievable effective capacity of the two users versus the delay exponent, with blocklength $n=400$, $\epsilon=10^{-6}$, and $\rho=[15\rm{dB},20dB]$. These curves show that the achievable effective capacity of both users decreases when the delay exponent becomes stringent. Interestingly, it is clear that when $\rho$ changes from 15dB to 20dB, the gain in the effective capacity of the strong user as compared to the one of the weak user is more significant. Specifically, when the delay exponent becomes stringent, the $\epsilon$ factor from the effective capacity formulation becomes more dominant, which renders the achievable capacity reach a minimum value.

\begin{figure}[ht]
\centering
  \includegraphics[width=0.9\linewidth]{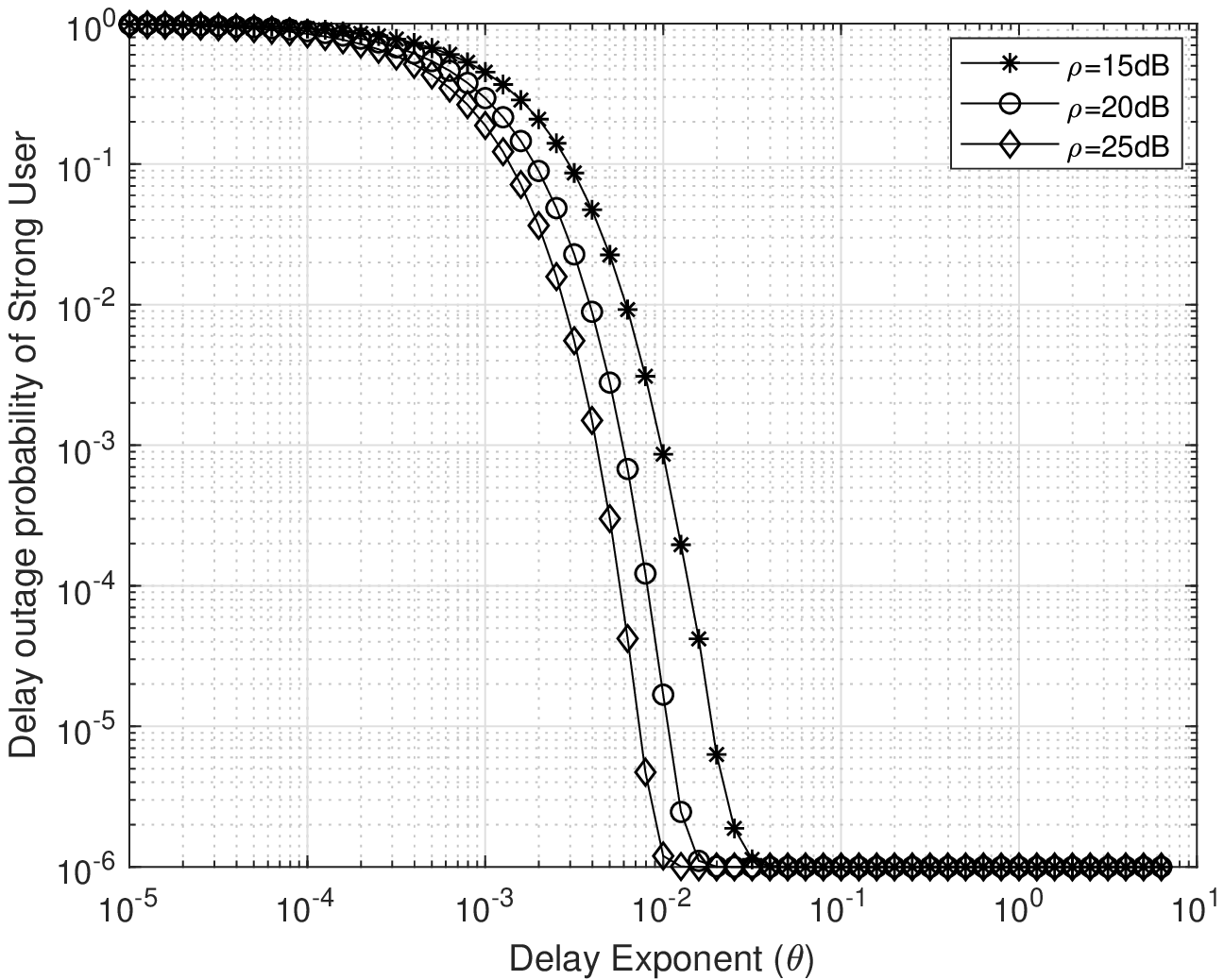}
\caption{Queuing delay violation probability of the strong user versus QoS exponent constraint ($\theta$), with $D_{{u}}^{\rm{max}}=400$, $\epsilon=10^{-6}$, and $n=400$.}
\label{fig:qdvp-theta-strong}
\end{figure}

\begin{figure}
\centering
 \includegraphics[width=0.9\linewidth]{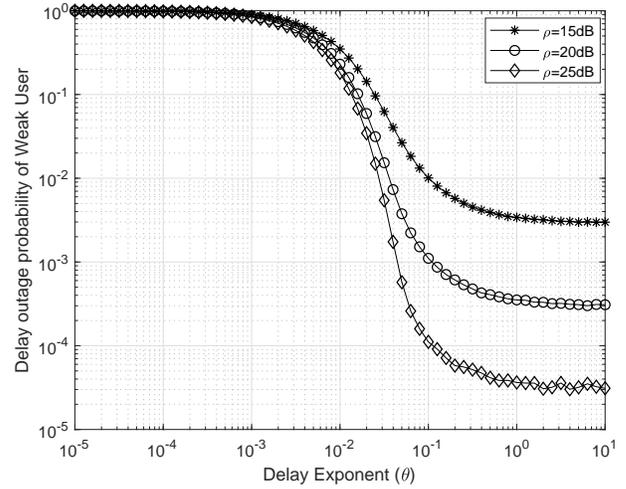}
\caption{Queuing delay violation probability of the weak user versus QoS exponent constraint ($\theta$), with $D_{{t}}^{\rm{max}}=400$,  $\epsilon=10^{-6}$, and $n=400$.}
\label{fig:qdvp-theta-weak}
\end{figure}

Fig. \ref{fig:qdvp-theta-strong} and Fig. \ref{fig:qdvp-theta-weak} show the plots of the queuing delay violation probability versus the delay exponent, for various values of $\rho$, namely, $\rho=[15\rm{dB},20dB,25dB]$, while considering strong and weak users. For these simulation results, $D_{{i}}^{\rm{max}}=400$, $\epsilon=10^{-6}$, and $n=400$. From the figure, it is  clear that when the delay exponent becomes stringent, the queuing delay violation probability does not improve below a certain value. As compared to the weak user, the strong user shows much improvement in queueing delay violation probability under QoS delay constraint. For both users, the queueing delay violation probability does not improve below a certain limit which is due to dominant factor of $\epsilon$ (short packet communications).
\section{Conclusion}
\label{sec:conclusion}
In this paper, the performance analysis of a downlink NOMA system in finite blocklength with effective capacity framework was studied in detail. The latency performance of the NOMA users with short packet communications was investigated by taking into account the queueing delay violation probability and the transmission error probability. Closed-form expressions for the achievable effective capacity of two paired users were derived, and their accuracy was confirmed using Monte-Carlo simulations. We also investigated the impact of the transmit signal-to-noise ratio and delay exponent on the achievable effective capacity and on the queueing delay violation probability. We showed that when the delay exponent becomes stringent, the queueing delay violation probability does not improve below a certain limit due to the dominant factor of the transmission error probability.

\end{document}